\documentclass[conference]{IEEEtran}

\usepackage[final]{changes}
\definechangesauthor[color = blue, name = {Mikael}]{MJ}
\definechangesauthor[color = green, name = {Arda}]{AA}

\usepackage{custompkg}
\usepackage{customdef}

\SetAlgoLined
\RestyleAlgo{ruled}

\author{\IEEEauthorblockN{Arda Aytekin and Mikael Johansson}
\IEEEauthorblockA{KTH Royal Institute of Technology\\
School of Electrical Engineering and Computer Science\\
Stockholm, SE-100 44\\
Emails: \texttt{\{aytekin,mikaelj\}@kth.se}}
}

\title{Harnessing the Power of Serverless Runtimes\\for Large-Scale
Optimization}

\begin{document}
\maketitle

\begin{abstract}
  \replaced[id=MJ]{The event-driven and elastic nature of serverless runtimes
  makes them a very}{Thanks to their event-driven and elastic nature, serverless
  runtimes are} efficient and cost-effective alternative\deleted[id=MJ]{s} for
  scaling up computations. So far, they have mostly been used for stateless,
  data parallel and ephemeral computations. In this work, we propose using
  serverless runtimes to solve generic, large-scale optimization problems.
  Specifically, we build a master-worker setup using AWS Lambda as the source of
  our workers, implement a \deleted[id=MJ]{synchronous} parallel optimization
  algorithm to solve a regularized logistic regression problem, and show that
  relative speedups up to 256 workers and efficiencies above 70\% up to 64
  workers can be expected. We also identify possible algorithmic and
  system-level bottlenecks, propose improvements, and discuss the limitations
  and challenges in realizing these improvements.
\end{abstract}


\section{Introduction}

\IEEEPARstart{D}{evelopments} in communication and data storage technologies
have made \replaced[id=MJ]{large-scale data collection}{the transmission and
storage of collected data} easier than ever. \replaced[id=MJ]{In order to
transform this data into insight or decisions, one often needs to solve a
large-scale optimization problem.}{In return, interesting optimization
problems are increasing in problem dimensions.}  \replaced[id=MJ]{Examples
include optimal classification and regression of data sets, such as, \eg,
those available in AWS Public Dataset Program~\cite{2019-AWSDataset}, and
training of deep neural networks for pattern recognition.}{For instance, wide
availability of hosted, big public datasets such as, \eg,
\cite{2019-AWSDataset}, as well as complex machine-learning models have led to
large-scale optimization problems.} Similarly, multi-stage stochastic
optimization problems \replaced[id=MJ]{appearing in finance, transportation
and energy domains also tend to have larger dimensions than what can}{, which
have applications in wide array of domains such as finance, transportation and
energy, also have large problem dimensions that cannot} be
\added[id=MJ]{reasonably} handled \deleted[id=MJ]{reasonably} by a single
computer.

Traditionally, large-scale problems have been tackled \replaced[id=MJ]{in}{with} high-performance
computing (HPC) environments. However, HPC environments are expensive and
inflexible in the sense that one has to deal with a lot of paperwork to apply
for computing power, write programs that obey certain rules and use \replaced[id=MJ]{specific}{certain}
libraries, estimate memory and running-time requirements of these programs and
submit them as jobs to the HPC environment accordingly, which are later
scheduled by the environment itself. These issues have led to HPC environments'
having limited reach (\cf\ the discussion in~\cite{2018-Shankar}).

Later, with the improvements \replaced[id=MJ]{in}{on the} virtualization technolog\replaced[id=MJ]{ies}{y}, cloud computing
providers started providing dedicated virtual machines (VMs) with different
memory and computing power \replaced[id=MJ]{configurations}{combinations} to their customers. Because these
dedicated VMs eliminate the burden of paperwork, provide customized programming
environments and do not involve job submissions, they have \replaced[id=MJ]{quickly gained wider adoption than}{better reach compared
to that of} HPC environments. However, these solutions still have relatively
coarse-grained resource combinations, which might be hard to choose from for a
given problem. Even though there exist works such as
Ernest~\cite{2016-Venkataraman} and Hemingway~\cite{2017-Pan} that help users
choose the correct resource combination for their problems, dedicated VMs are
still hard to rescale for new, differently-sized problems. In addition, needing
to provision VMs and pay for their idle times are among the reasons that limit
VMs' reach in many scientific applications.

More recently, \replaced[id=MJ]{improvements in}{the improvements on the} container-based solutions have opened an
alternative path: serverless runtimes~\cite{2018-Lambda,2018-AFunctions,
2018-GFunctions}. Serverless runtimes are compute services that let users run
their programs in isolated containers \emph{without} the need for managing or
provisioning servers. The main motivation behind the serverless runtimes is, for
the providers, that they can simply \replaced[id=MJ]{offer current excess capacity}{provide the available computing resources} at
their backends temporarily to the customers. From the end-user\deleted[id=MJ]{'s} perspective,
serverless runtimes are advantageous thanks to their \emph{event-driven} and
\emph{elastic} nature. Users can not only activate serverless runtimes based on
events such as HTTP requests, and thus, only pay for what they use, but also
change the resource allocation and number of workers of the runtimes dynamically
as needed. Moreover, all major providers support custom programming runtimes,
and give away free usage every month. As an example, users get, from each
provider, roughly 9 hours of free computing time every month, should they choose
to allocate 128 MB memory for their runtimes and spawn 100
workers\footnote{At the time of writing, major providers are giving $400\,000$
GB-seconds of free computing time every month.}.

This elasticity of serverless runtimes, however, comes at the expense of some
limitations. First, serverless runtimes are \emph{stateless}. As such, users of
serverless runtimes are responsible for keeping track of the states properly in
stateful application domains, \ie, when, for instance, running scientific
computations or solving optimization problems. Second, serverless runtimes are
designed for event-driven, ephemeral applications such as, \eg, manipulating a
database record upon an HTTP request, which leads to \emph{limited} computation
times and memory. As a result, careful design of the application code is needed
to stay strictly within the bounds, or handle dynamic joining of new workers and
leaving of dying workers in long-lived applications.  Due to the dynamic nature
of the communications, \deleted[id=MJ]{mature} message-passing libraries such as
MPI that require static network of nodes are not a viable option, either.

Even though serverless runtimes have been around since the late 2014, \added[id=MJ]{the} literature
involving the serverless runtimes and scientific computations is relatively
sparse. There exist studies that examine the task completion
latencies~\cite{2018-Gorlatova} and optimize the price of running
applications~\cite{2018-Elgamal} in serverless runtimes. Works such as
\cite{2017-Ishakian,2018-GCSMLModel} propose using serverless runtimes to serve
(already trained) deep-learning models. However, these applications do not
require persistent states or involve frequent communications among workers of
the serverless runtimes. In that regard, studies that implement linear algebra
primitives~\cite{2018-Shankar} and do hyperparameter optimization on neural
networks~\cite{2018-Feng} are more relevant examples to using serverless
runtimes for stateful and long-lived applications.

In this paper, we try to assess opportunities, limitations and challenges of the
serverless runtimes for solving \emph{generic}, large-scale optimization
problems. More specifically,
\begin{itemize}
  \item We propose using serverless workers for solving optimization problems,
    \emph{collectively}, in a distributed way. This is, to the best of our
    knowledge, the first time serverless runtimes have been used for this
    purpose. In contrast to the hyperparameter optimization~\cite{2018-Feng},
    which can be carried out by each worker \emph{independently} of each other,
    our work requires that workers share their states at each iteration during
    the lifetime of the program.
  \item We show that relative speedups could be expected up to 256 workers and
    efficiencies above 70\% could be expected up to 64 workers, even in a naive
    implementation.
  \item We identify possible algorithmic and system-level bottlenecks, propose
    improvements, and point out the limitations and challenges in implementing
    the improvements.
\end{itemize}
The rest of the paper is organized as follows. In
Section~\ref{sec:serverless-motivation}, we motivate our work, introduce our
experimental setup and list our goals. In
Section~\ref{sec:serverless-experiments}, we briefly \replaced[id=MJ]{describe the}{mention about out}
experiments, and in Section~\ref{sec:serverless-results}, we \replaced[id=MJ]{report}{give} our results.
Finally, in Section~\ref{sec:serverless-conclusion}, we \replaced[id=MJ]{summarize our findings}{finalize our paper} and
discuss about possible \replaced[id=MJ]{ways to improve scalability even further.}{improvements to obtain better scalability.}

\section{Motivation}\label{sec:serverless-motivation}

\begin{figure*}
  \centering
  \includegraphics{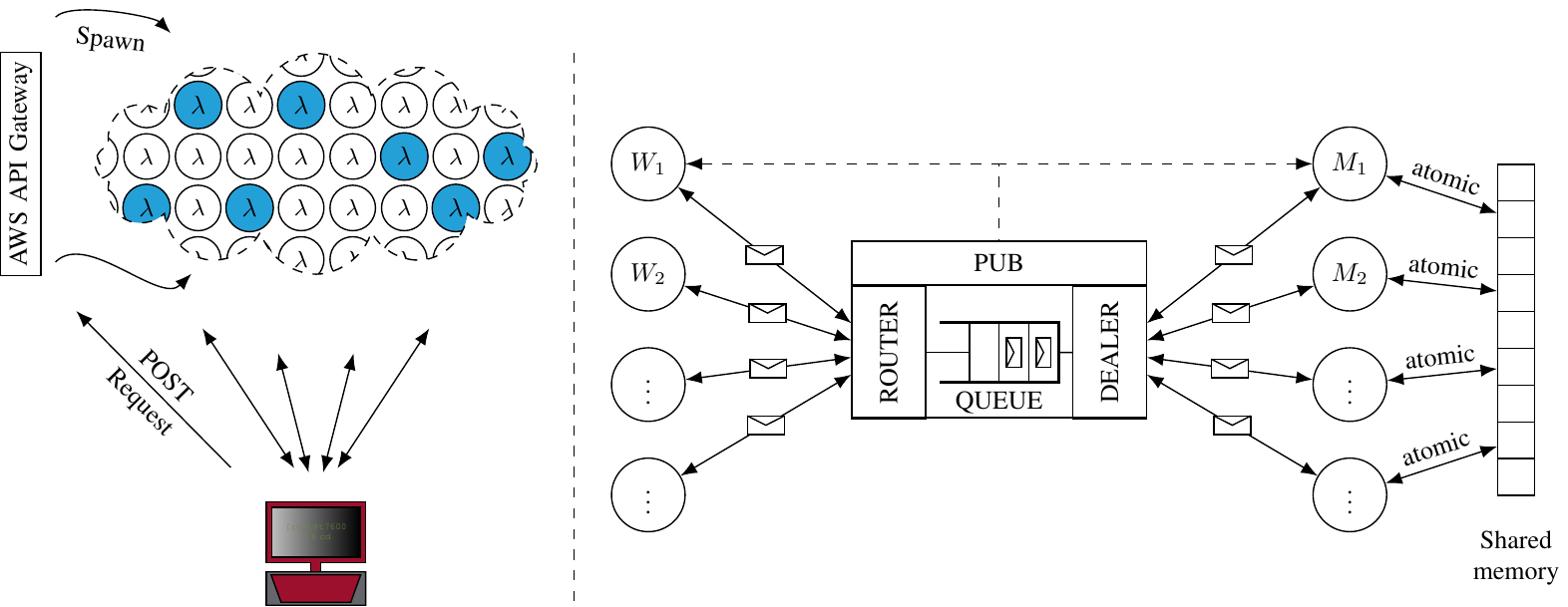}
  \caption{The external (left) and internal (right) views of the setup. The
    publisher (PUB) socket broadcasts important signals, such as the termination
    signal, to the workers and masters.}\label{fig:serverless-setup}
\end{figure*}

To \deleted[id=MJ]{empirically}evaluate the performance and \deleted[id=MJ]{understand} possible limitations of
the serverless runtimes for distributed optimization problems, we choose to
focus on the problems of the form
\begin{align}\label{eqn:serverless-opt-prob}
  \begin{aligned}
    & \minimize_{x \in \set{R}^{d}}
    & & \funcit[\phi]{x} \coloneqq \sum_{n = 1}^{N} \funcit[f_{n}]{x} +
      \funcit[h]{x} \,.
  \end{aligned}
\end{align}
\replaced[id=MJ]{This loss function}{where the total loss function,} $\funcit[\phi]{x}$, \added[id=MJ]{which appears in many applications, }consists of two main parts.
The first part is a sum of $N$ smooth functions, $\funcit[f_{n}]{x}$,  and the
second part is a possibly non-smooth function, $\funcit[h]{x}$, of the $d$-dimensional decision vector $x$.

Many data-driven machine-learning problems and multi-stage stochastic decision
problems can be cast as optimization problems on the
form~\eqref{eqn:serverless-opt-prob}. In the case of machine-learning problems,
the smooth part encodes the data loss while the non-smooth part is generally
used to give preference to a particular solution with desirable properties. For
instance, given a data matrix $A$ and a vector $b$, in ridge regression
problems, one tries to solve
\begin{align*}
  \begin{aligned}
    & \minimize_{x \in \set{R}^{d}}
    & & \sum_{n = 1}^{N} \norm{\dotprod{a_{n}}{x} - b_{n}}_{2}^{2} +
      \frac{\lambda_{2}}{2} \norm{x}_{2}^{2} \,,
  \end{aligned}
\end{align*}
where the first part is the sum of squared residuals coming from ordinary
least-squares and the second part is used to prefer solutions with smaller
norms. Similarly, in $\ell_{1}$-penalized logistic regression problems, one
tries to solve
\begin{align*}
  \begin{aligned}
    & \minimize_{x \in \set{R}^{d}}
    & & \sum_{n = 1}^{N} \func[log]{1 + \func[exp]{-b_{n}\dotprod{a_{n}}{x}}} +
      \lambda_{1} \norm{x}_{1}^{1} \,,
  \end{aligned}
\end{align*}
where $b_{n} \in \lbrace -1, 1 \rbrace$ is the (binary) label for each of the
$N$ samples, and the second part in the total loss is used to promote sparse
solutions. In the case of multi-stage stochastic optimization problems
\begin{align*}
  \begin{aligned}
    & \minimize_{x \in \set{R}^{d}}
    & & \sum_{n = 1}^{N} \pi_{n} \funcit[f_{n}]{x} + \indicator{Mx} \,,
  \end{aligned}
\end{align*}
where $\pi_{n}$ denotes the probability of occurring of a scenario in a scenario
tree and $\indicator{\cdot}$ denotes the indicator function of a set, $\set{X}$,
the first part encodes the expected cost of the scenario tree whereas the second
part encodes the nonanticipativity of stage variables~\cite{2018-Eckstein}.

One approach to solving problems of the form~\eqref{eqn:serverless-opt-prob} is
to use \emph{proximal algorithms}~\cite{2014-Parikh}. Different proximal
algorithms iteratively use the proximal operator
\begin{align*}
  x_{k+1} & = \func[prox_{\gamma_{k}\operatorname{\mathit{f}}}]{x_{k}} \\
          & = \argmin[x]{\funcit{x} + \frac{1}{2\gamma_{k}} \norm{x -
          x_{k}}_{2}^{2}}
\end{align*}
for some function $\funcit{\cdot}$ and step size $\gamma_{k}$. When the loss
function is naturally split into two parts, as
in~\eqref{eqn:serverless-opt-prob}, where both parts are closed and proper
convex functions, one common approach is to use the \emph{alternating direction
method of multipliers} (ADMM), which, at each iteration $k$, solves the
following subproblems~\cite{2010-Boyd}:
\begin{align}
  x_{k+1} & = \argmin[x]{\sum_{n=1}^{N} \funcit[f_{n}]{x} + \frac{\rho_{k}}{2}
    \norm{x - \left(z_{k} - u_{k}\right)}_{2}^{2}}
    \label{eqn:serverless-admm-begin} \\
  z_{k+1} & = \argmin[z]{\funcit[h]{z} + \frac{\rho_{k}}{2} \norm{z -
    \left(x_{k+1} + u_{k}\right)}_{2}^{2}} \\
  u_{k+1} & = u_{k} + x_{k+1} - z_{k+1} \,. \label{eqn:serverless-admm-end}
\end{align}
Basically, ADMM iterations alternate between $x$ and $z$-updates by using the
proximal operators of the smooth and non-smooth parts of the \added[id=MJ]{augmented Lagrangian of the} loss function,
respectively, and then update the dual variable $u$. Here, $\rho_{k} > 0$ is
called the penalty parameter.

ADMM is particularly useful when each part of the loss is proximable, \ie, the
proximal operator for each part is efficiently obtained, whereas the total sum
is not~\cite{2014-Parikh}. Moreover, since ADMM handles each part separately,
this method is suitable for distributed-memory architectures such as, \eg, the
master-worker setups, where worker nodes hold chunks of the smooth loss, and the
master node keeps track of the common decision vector and is responsible for
handling the non-smooth loss. However, in these setups, one needs to rewrite the
ADMM iterations
\eqref{eqn:serverless-admm-begin}--\eqref{eqn:serverless-admm-end} to obtain the
so-called global variable consensus formulation~\cite{2010-Boyd}:
\begin{align}
  x_{k+1}^{w} & = \argmin[x]{\sum_{n \in \set{N}_{w}} \funcit[f_{n}]{x} +
    \frac{\rho_{k}}{2} \norm{x - \left(z_{k} - u_{k}^{w}\right)}_{2}^{2}} \quad
    \forall w \label{eqn:serverless-consen-begin} \\
  z_{k+1} & = \argmin[z]{\funcit[h]{z} + \frac{N\rho_{k}}{2} \norm{z -
    \left(\bar{x}_{k+1} + \bar{u}_{k}\right)}_{2}^{2}}
    \label{eqn:serverless-consen-mid} \\
  u_{k+1}^{w} & = u_{k}^{w} + x_{k+1}^{w} - z_{k+1} \quad \forall w \,.
    \label{eqn:serverless-consen-end}
\end{align}
Here, each worker $w$ updates its own copy of $x^{w}$ using its local dataset
$\set{N}_{w}$, and the master updates the global variable $z$ using the averaged
variables ($\bar{x}$ and $\bar{u}$) \replaced[id=MJ]{of}{coming from} workers.

\subsection{Setup}

In this paper, we construct a master-worker setup similar to those discussed
in~\cite{2013-Li,2017-Xiao}, and use AWS Lambda as the source of our worker
nodes. Currently, AWS Lambda does not allow for inbound network connections.
Hence, one cannot obtain a fully connected network of master and worker nodes.
For this reason, we build a star network, and assign a \replaced[id=MJ]{local}{32-core} server, \ie,
\emph{the scheduler}, as the central node (see
Figure~\ref{fig:serverless-setup}). Each node in the star network is connected
to the central node with a point-to-point connection. We use
\texttt{ZMQ}~\cite{2017-ZMQ} to handle dynamic joining and leaving of workers in
the network, \texttt{cereal}~\cite{2018-USCiLab} to serialize and deserialize
data, and \texttt{cURL}~\cite{2018-curl} and AWS API Gateway to spawn AWS Lambda
functions.

In this setup, the scheduler is responsible for spawning and orchestrating
masters and workers. Given a fixed-size problem, the scheduler generates POST
requests for \added[id=MJ]{the} AWS API Gateway to spawn $W$ workers, and embeds the necessary
states of the algorithm such as, \eg, \deleted[id=MJ]{the} problem information and \deleted[id=MJ]{the} local
solver options, inside the requests. It uses \texttt{ZMQ}'s router socket to
fair-queue messages coming from the workers. To alleviate the delays in
processing the message queue, the scheduler spawns one local master thread per
$\bar{W}$ workers, and uses \texttt{ZMQ}'s dealer socket to distribute the
messages to the master threads in a round-robin fashion. Master threads process
the queue in parallel, average the local variables of the workers using atomic
operations, and finally update the global $z$ variable (\cf\
Equation~\eqref{eqn:serverless-consen-mid}). If the \replaced[id=MJ]{stopping}{termination} criterion is
satisfied, the scheduler sends a termination signal to the workers and the
masters. Otherwise, it broadcasts the new penalty parameter along with the
updated $z$ variable to the workers. \replaced[id=MJ]{Pseudocode}{We list the pseudocode} for the scheduler's
logic \added[id=MJ]{is listed} in Algorithm~\ref{alg:serverless-scheduler}.

\begin{algorithm}[ht]
  \LinesNumbered
  \SetKwFunction{Receive}{receive}
  \SetKwFunction{Broadcast}{broadcast}
  \SetKwFunction{Converged}{converged}
  \SetKwFunction{Spawn}{spawn}
  \SetKwFunction{Penalty}{new\_penalty}

  \KwIn{Total number of samples, $N$; problem dimension, $d$; density of
    non-zero features for each sample, $p$; possibly non-smooth function,
    $\funcit[h]{\cdot}$; number of workers, $W$; maximum number of workers per
    master, $\bar{W}$; initial penalty parameter, $\rho_{0}$; maximum number of
    ADMM iterations, $K$; primal residual tolerance, $\epsilon_{r}$; dual
    residual tolerance, $\epsilon_{s}$; minimum number of local solver
    iterations, $K_{w}$; gradient norm tolerance, $\epsilon_{g}$; relative
    function value improvement $\epsilon_{f}$.}

  \KwOut{Optimizer, $x^{\star}$.}

  \ForEach(API Gateway calls){$w = 1$ \KwTo $W$}{
    $N_{w} \gets N/W$\;
    \Spawn{$w$; $N_{w}$, $d$, $p$, $\rho_{0}$, $K_{w}$, $\epsilon_{g}$,
    $\epsilon_{f}$}\;
  }

  Initialize $k \gets 0$, $t \gets 0$, $\omega \gets 0$, $z_{k} \gets 0$ and
  $z_{k+1} \gets 0$\;

  \ForEach(in parallel){$m = 1$ \KwTo $\ceil{W / \bar{W}}$}{
    \Repeat{\texttt{TERM} is received}{
      \Receive{$q_{k}^{w}$, $\omega_{k+1}^{w}$}\;
      $q \gets q + q_{k}^{w}$\tcp*[r]{(atomic) reduce}
      $\omega \gets \omega + \omega_{k+1}^{w} / N$\tcp*[r]{(atomic) reduce}
      \If(block others){All workers have returned}{
        \tcp{Update the global $z$}
        $z_{k} \gets z_{k+1}$\;
        $z_{k+1} \gets \argmin[z]{\funcit[h]{z} + \frac{N\rho_{k}}{2} \norm{z -
          \omega}_{2}^{2}}$\label{alg:serverless-soft-thres}\;
        $r \gets \sqrt{q}$\;
        $s \gets \rho_{k} \norm{z_{k+1} - z_{k}}_{2}$\;
        $\rho_{k+1} \gets $ \Penalty{$\rho_{k}$, $r$, $s$}\;
        \eIf{\Converged{$r$, $s$, $k$; $\epsilon_{r}$, $\epsilon_{s}$, $K$}}{
          Signal \texttt{TERM}\;
          $x^{\star} \gets z_{k+1}$\;
        }{
          \Broadcast{$\rho_{k+1}$, $z_{k+1}$}\;
          $q \gets 0$, $\omega \gets 0$, $k \gets k + 1$\;
        }
      }
    }
  }
  \Return{$x^{\star}$}\;
  \caption{Scheduler logic.}\label{alg:serverless-scheduler}
\end{algorithm}

Worker nodes, on the other hand, load their local problem data and initialize
their local solvers based on the state information they receive in the POST
requests. Local problem data is not present in the scheduler; instead, the
scheduler simply provides enough information so that the workers could either
fetch a batch of data samples from hosted datasets or generate the problem data
from its closed-form formulation. Then, they update their local primal and dual
variables (\cf\ Equations~\eqref{eqn:serverless-consen-begin} and
\eqref{eqn:serverless-consen-end}) with the penalty parameter and global $z$
variable they receive from the scheduler, and send back the updated ones. We
list the pseudocode in Algorithm~\ref{alg:serverless-worker}.

\begin{algorithm}[ht]
  \LinesNumbered
  \SetKwFunction{Receive}{receive}
  \SetKwFunction{Send}{send}

  \KwIn{$N_{w}$, $d$, $p$, $\rho_{0}$, $K_{w}$, $\epsilon_{g}$, $\epsilon_{f}$.}

  Load problem data with $N_{w}$, $d$ and $p$
  (Section~\ref{sec:serverless-experiments})\;

  Initialize local solver with $K_{w}$, $\epsilon_{g}$ and $\epsilon_{f}$
  (Section~\ref{sec:serverless-experiments})\;

  Initialize $x_{0}$, $k \gets 0$, $u_{0} \gets 0$ and $z_{0} \gets 0$\;

  \Repeat{\texttt{TERM} is received}{
    $r_{k} \gets x_{k} - z_{k}$\;
    $u_{k+1} \gets u_{k} + r_{k}$\;
    $x_{k+1} \gets \argmin{\sum_{n = 1}^{N_{w}} \funcit[f_{n}]{x} +
      \frac{\rho_{k}}{2} \norm{x - z_{k} +
      u_{k+1}}_{2}^{2}}$\label{alg:serverless-fista}\;
    $q_{k}^{w} \gets \norm{r_{k}}_{2}^{2}$\;
    $\omega_{k+1}^{w} \gets x_{k+1} + u_{k+1}$\;
    \Send{$q_{k}^{w}$, $\omega_{k+1}^{w}$}\;
    \Receive{$\rho_{k+1}$, $z_{k+1}$}\;
    $k \gets k + 1$\;
  }
  \caption{Worker logic.}\label{alg:serverless-worker}
\end{algorithm}

\subsection{Goals}

Our goals in this paper are to assess the performance of serverless runtimes
when solving generic optimization problems, and identify possible bottlenecks as
well as challenges in addressing them. To this end, we measure the following.

\textbf{Relative speedup and efficiency.} Perhaps the most important measures
when evaluating the performance of parallel computations are the relative
speedup and efficiency. Relative speedup is the speedup obtained in the new
architecture with respect to the old one, \ie, $S_{\text{new}} = t_{\text{old}}
/ t_{\text{new}}$, where $t_{\text{old}}$ and $t_{\text{new}}$ are the
wall-clock times of finishing a task in the old and new architecture,
respectively. The efficiency of the new architecture gives an indication of how
well the resources are utilized, and is defined as $E_{\text{new}} =
S_{\text{new}} / (W_{\text{new}} / W_{\text{old}})$, where $W_{\text{old}}$ and
$W_{\text{new}}$ are the number of workers employed in the old and new
architecture, respectively.

\begin{figure}
  \centering
  \includegraphics[width = \linewidth]{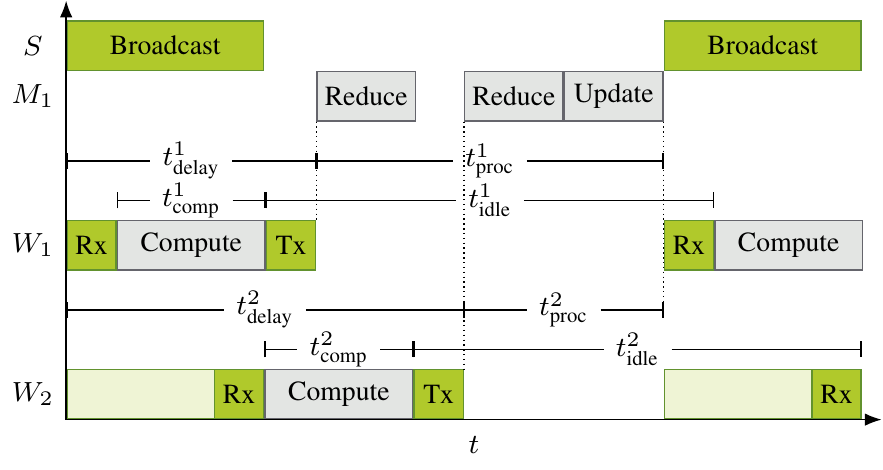}
  \caption{Sample timing diagram for the scheduler ($S$), master ($M_{1}$) and
    workers ($W_{1}$, $W_{2}$) in a 1-master-2-worker setup. ``Rx'' and ``Tx''
    stand for ``reception'' and ``transmission,''
    respectively.}\label{fig:serverless-utilization}
\end{figure}

\textbf{Utilization.} To identify bottlenecks in both the algorithm and our
experimental setup, we want to understand how the worker functions use their
time. To this end, we measure three major utilization metrics: \emph{idle time},
\emph{computation time} and \emph{delay time} of each worker (see
Figure~\ref{fig:serverless-utilization}). The \emph{idle time} of a worker is
measured from the time it sends its local $x$ and $u$ variables until the time
it receives the updated $z$ variable from the scheduler. Thus, the idle time
includes both the total \emph{communication time} for the variables and the
\emph{processing time} at the scheduler, \ie, $t^{w}_{\text{idle}} =
t^{w}_{\text{comm}} + t^{w}_{\text{proc}}$. The worker's \emph{computation
time}, $t^{w}_{\text{comp}}$, is the time from when it receives a new global $z$
variable from the scheduler until it returns its updated local variables. Both
idle time and computation times are measured by the worker itself. Finally, the
\emph{delay time} associated with a worker as observed from a master is the time
between when the scheduler broadcasts the global $z$ variable until the master
starts processing the corresponding worker's $x$ and $u$ updates. The delay time
includes both the total communication time and the computation time of the
worker, \ie, $t^{w}_{\text{delay}} = t^{w}_{\text{comm}} + t^{w}_{\text{comp}}$.
From these metrics, we compute the total communication time of a worker as
$t^{w}_{\text{comm}} = t^{w}_{\text{delay}} - t^{w}_{\text{comp}}$. Similarly,
we estimate the effect of \emph{queuing} at the scheduler node by subtracting
the delay time from the idle time of the workers, \ie, $t^{w}_{\text{idle}} -
t^{w}_{\text{delay}} = t^{w}_{\text{proc}} - t^{w}_{\text{comp}}$. Ideally,
processing times at the scheduler should not exceed the workers' computation
times.

\textbf{Cold starting.} Worker functions are not only limited in computation
time per invocation but are also stateless. Hence, serverless runtimes suffer
from the cold-starting phenomenon, which is defined as the penalty of getting
serverless code ready to run~\cite{2017-Baldini}. We measure cold-starting times
of the workers from the time the scheduler generates the API request until the
time the worker contacts the scheduler for the first time. Cold-starting times
include transmission of the API request, spawning of the worker and the loading
of local data at the worker. Because, in long-lived optimization algorithms, the
scheduler needs to spawn new workers to replace those approaching their time
limits, cold starting of workers should be small relative to the computation
time of the workers.

\textbf{Responsiveness.} The last measure we are interested in assessing is how
fast the worker functions respond to the scheduler at each iteration. Because
serverless functions are isolated containers that share memory, CPU and network
resources with others in the service provider's backend, their response times
can get perturbed by the actual load of the corresponding node in the backend.
We would like to understand if there are any \emph{stragglers} that consistently
fall behind the rest.

\section{Experiments}\label{sec:serverless-experiments}

In our experiments, we follow the procedure outlined in~\cite{2007-Koh} to
generate a random instance of $\ell_{1}$-penalized logistic regression problem:
\begin{align*}
  \begin{aligned}
    & \minimize_{x \in \set{R}^{d}}
    & & \sum_{n = 1}^{N} \func[log]{1 + \func[exp]{-b_{n}\dotprod{a_{n}}{x}}} +
    \lambda_{1} \norm{x}_{1}^{1}
  \end{aligned}
\end{align*}
with $N = 600\,000$ samples, $d = 10\,000$ features, $p = 0.001$ density\replaced[id=MJ]{ (proportion of non-zero features in each sample)}{, \ie,
density of non-zero features for each sample}, and $\lambda_{1} = 1$. Every
sample has equal probability of having a positive (or negative) label, \ie,
$\prob{b_{n} = 1} = \prob{b_{n} = -1} = 0.5$. Indices of the non-zero features
for each sample are selected uniformly at random without replacement, whereas
the values of non-zero features for samples with positive (or negative) label
are drawn independently from a normal distribution $\func[\mathcal{N}]{\nu, 1}$,
where $\nu$ is drawn from a uniform distribution on $[0,1]$ (or $[-1,0]$).

For $\ell_{1}$-penalized logistic regression problems, the solution to the
subproblem on Line~\ref{alg:serverless-soft-thres} of
Algorithm~\ref{alg:serverless-scheduler} can be obtained easily by applying the
soft thresholding operator
\begin{align*}
  \func[\mathcal{S}]{a; b} = \func[max]{0, \left(1 - b / \abs{a}\right)} a
\end{align*}
element-wise to $\omega$ with $b = \lambda_{1} / (N\rho_{k})$. However, the
subproblem on Line~\ref{alg:serverless-fista} of
Algorithm~\ref{alg:serverless-worker} does not have a closed-form solution.
Hence, we \deleted[id=MJ]{approximately} solve \replaced[id=MJ]{this subproblem (approximately)}{the subproblem by} using an iterative method,
FISTA~\cite{2009-Beck}\replaced[id=MJ]{ with backtracking.}{, with a backtracking procedure.} As\deleted[id=MJ]{the} termination
criterion\added{ for FISTA}, we choose to \replaced[id=MJ]{require that}{have} either $\norm{g_{\tilde{k}}}_{2} \leq \epsilon_{g}
= 10^{-2}$ or $(f_{\tilde{k}-1} - f_{\tilde{k}}) / f_{\tilde{k}-1} \leq
\epsilon_{f} = 10^{-12}$, where $g_{\tilde{k}}$ and $f_{\tilde{k}}$ are the
gradient and function value of the augmented loss at (inner) iteration
$\tilde{k}$, respectively. We observe that the gradient norm tolerance and
relative function value improvement criteria lead to different number of (inner)
iterations for different subproblems, and thus, nonuniform load distributions on
the workers. \replaced[id=MJ]{To observe any external effects on the load of workers, we therefore perform}{As a result, we choose to have} two sets of experiments by forcing
FISTA to run at least $K_{w} = 1$ (nonuniform load) and $K_{w} = 50$ iterations
(uniform load)\deleted[id=MJ]{, to observe any external effects on the load distribution of the
workers}.

Although problem instances with the aforementioned dimensions can fit in a
single AWS Lambda worker with 128 MB of memory, they are too large to handle
with $W = 1$ or $W = 2$ workers within the computation time limit of 15
minutes\footnote{In fact, increasing the memory size of AWS Lambda functions
also improves their CPU and network shares, which helps with the computation
time. However, one can still construct a large enough problem that cannot be
handled by fewer than four workers regardless of their CPU shares.}.  As a
result, we start with spawning $W = 4$ workers and double the number of workers
until we do not observe further relative speedup.  We consider the ADMM
iterations as converged when either both primal and dual residual norms are
small enough, \ie, $r \leq \epsilon_{r} = 2 \cdot 10^{-2}$ and $s \leq
\epsilon_{s} = 2 \cdot 10^{-2}$, or $K = 100$ iterations have passed. Finally,
we use the following rule~\cite{2010-Boyd} to adjust the new penalty parameter
at each iteration:
\begin{align*}
  \rho_{k+1} = \begin{cases}
    2\rho_{k}   & \text{if } r > 10s \,, \\
    0.5\rho_{k} & \text{if } s > 10r \,, \\
    \rho_{k}    & \text{otherwise,}
  \end{cases}
\end{align*}
starting with $\rho_{0} = 1$.

\section{Results}\label{sec:serverless-results}

\begin{figure}
  \centering
  \includegraphics{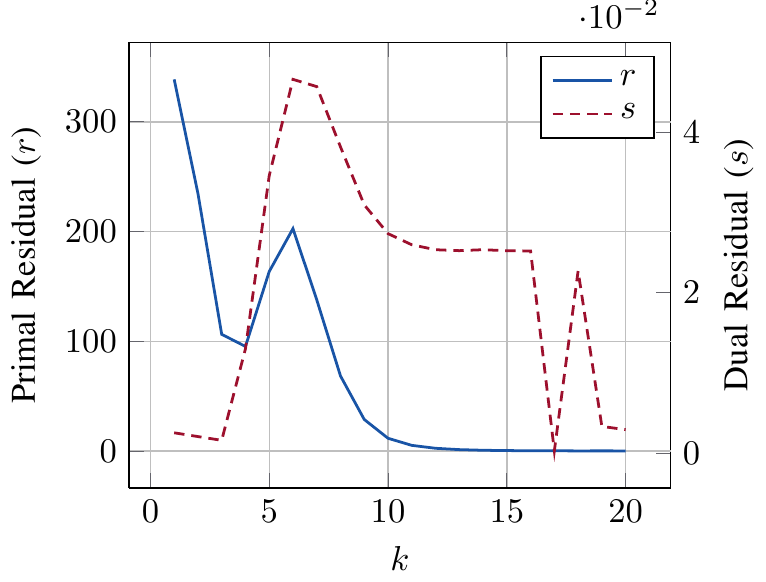}
  \caption{Convergence of the residuals for $W = 64$ workers and $K_{w} =
    1$.}\label{fig:serverless-convergence}
\end{figure}

In our experiments, we observe speedups in wall-clock times of ADMM iterations
up to $W = 256$ workers. In all the experiments, we observe that ADMM iterations
converge within at most $K = 23$ iterations by satisfying the primal and dual
residual tolerance values. In Figure~\ref{fig:serverless-convergence}, we
provide traces of the residuals for $W = 64$ workers when the workers had
nonuniform load distributions.

\begin{figure*}
  \centering
  \includegraphics{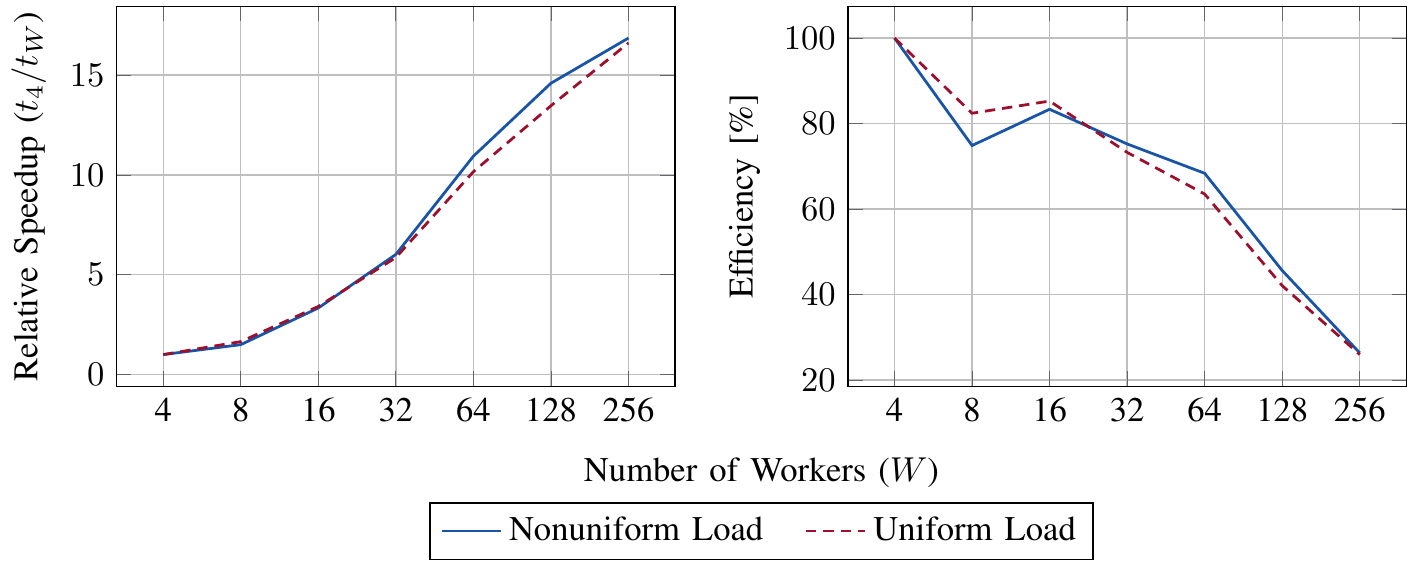}
  \caption{Speedup and effiency plots. Speedup and efficiency of the algorithm
    are reported relative to $W = 4$ workers, since fewer workers are not
    sufficient to solve the problem within the time limit.}
  \label{fig:serverless-speedup}
\end{figure*}

\textbf{Relative speedup and efficiency.} Because our problem instances cannot
be solved by fewer than four workers, we report relative speedup and efficiency
metrics with respect to $W = 4$. In Figure~\ref{fig:serverless-speedup}, we
observe that relative speedups up to 17 times can be expected in both uniform
and nonuniform load\deleted[id=MJ]{ing} scenarios, which translates to 26\% efficiency.

\begin{figure}
  \centering
  \includegraphics{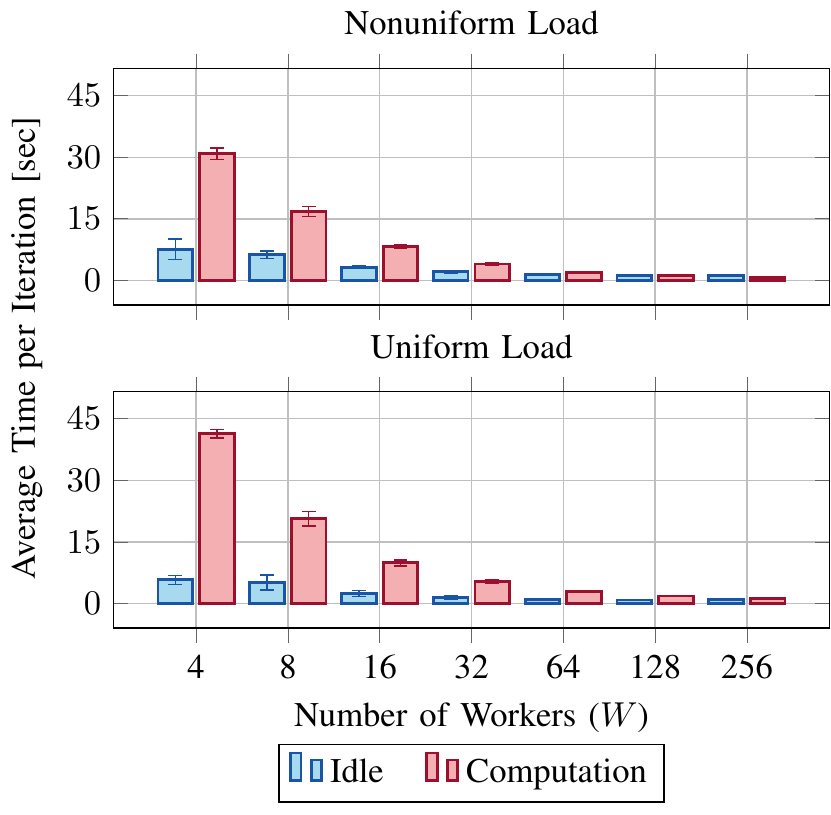}
  \caption{Average utilization of Lambda workers. Time spent in idle and
    computation is first averaged over total number of iterations, and then over
    total number of Lambda functions. Error bars show one standard deviation
    among the workers.}\label{fig:serverless-average}
\end{figure}

We also observe that there is a sharp decrease in efficiency when going from $W
= 64$ (74\%) to $W = 256$ (26\%) workers. This is best explained in
Figure~\ref{fig:serverless-average}, which shows the average idle and
computation times per iteration. As can be seen, after $W = 64$ workers, average
idle time starts beating average computation time. Basically, when we increase
the number of workers, average computation time constantly decreases. On the
contrary, average idle time decreases up to a point and then increases again
with increasing number of workers. The reason is that increasing the number of
workers improves the worst-case solution times of subproblems \added[id=MJ]{(which get smaller as the number of workers increases)}, which in turn
improves the idle time up to the level set by transmission time of the decision
vector. After this level, queuing effects take over with increasing number of
workers. In the ideal case, instead of fixing the problem size and increasing
the number of workers, one should aim at increasing both the problem size and
the number of workers to benefit from more computing power, which is in line
with Gustafson's Law~\cite{1988-Gustafson}.

The main difference between uniform and nonuniform loads is that the average
computation\deleted[id=MJ]{s} times are increased per iteration and the variance in both idle and
computation times is decreased for uniform loads. This is because we make the
local solvers run for roughly the same number of iterations ($K_{w} = 50$).

\begin{figure*}
  \centering
  \includegraphics{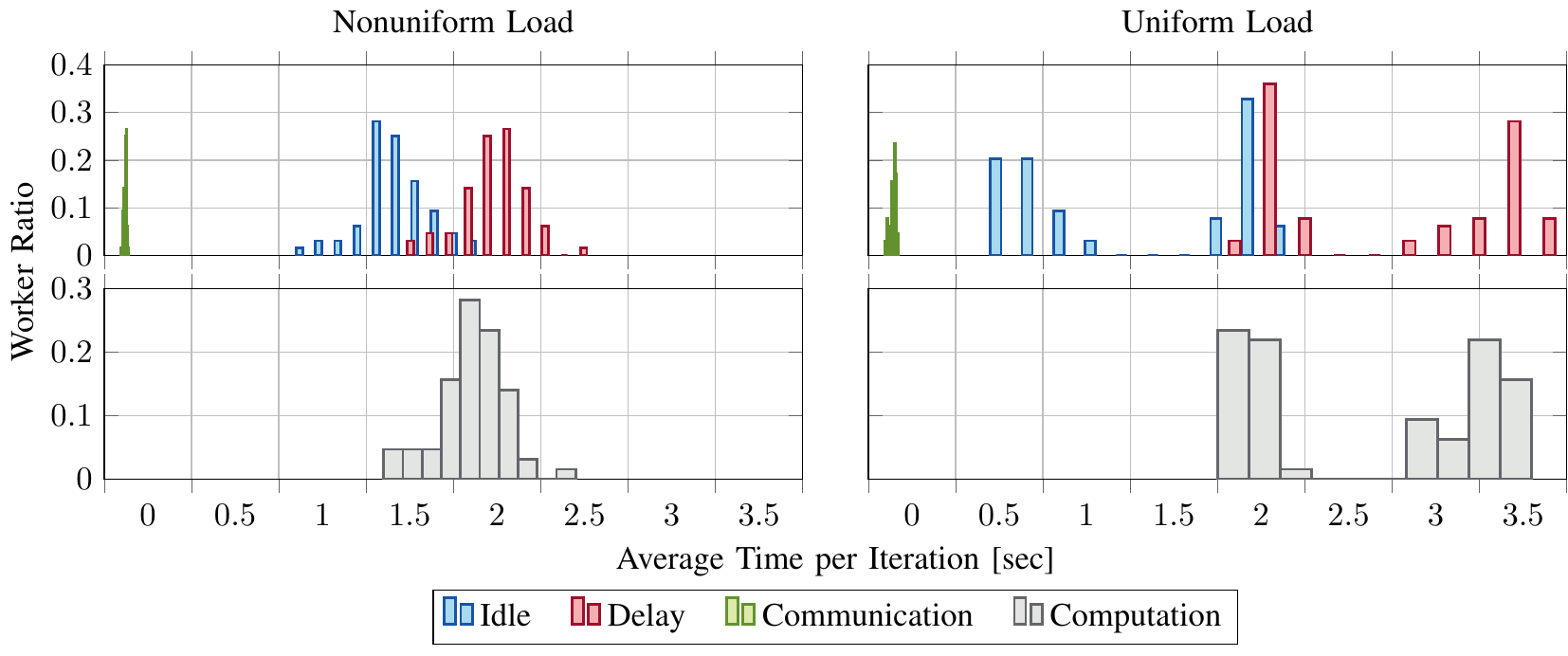}
  \caption{Utilization histograms for Lambda workers when $W = 64$ workers are
  used.}\label{fig:serverless-histograms}
\end{figure*}

\textbf{Utilization.} Even though AWS Lambda does not guarantee any performance
measures other than the built-in fault tolerance and allocation of CPU power,
network bandwidth and disk input/output proportional to the selected memory size
of the workers, we have observed consistent behavior in workers' performance
during our experiments (see Figure~\ref{fig:serverless-histograms} for a sample
histogram for $W = 64$ workers).

\begin{figure*}
  \centering
  \includegraphics{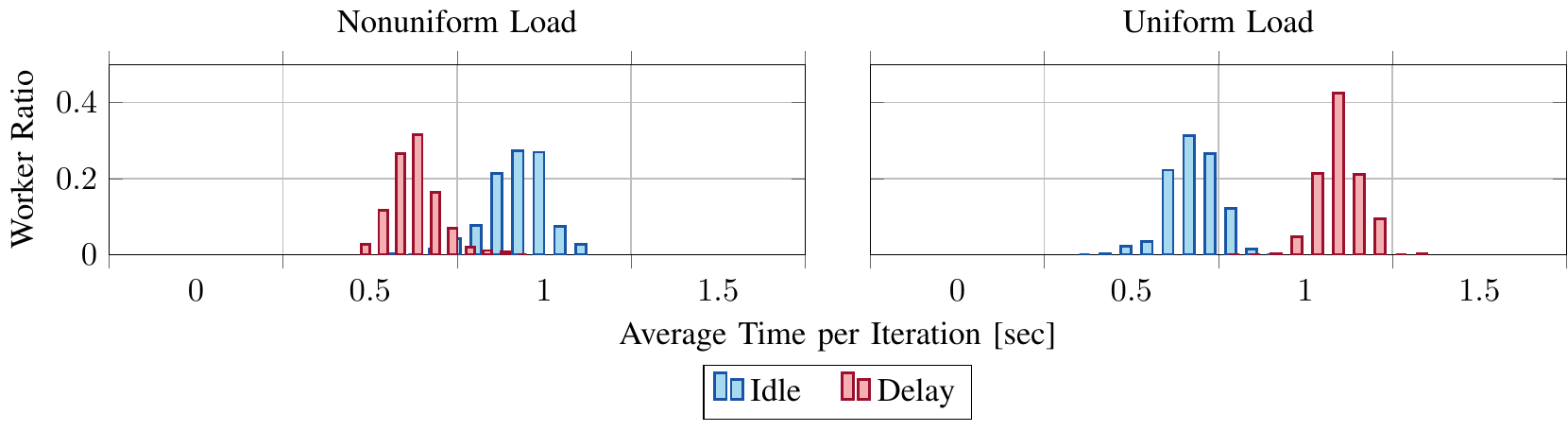}
  \caption{Utilization histograms for Lambda workers when $W = 256$ workers are
    used.}\label{fig:serverless-queue}
\end{figure*}

As expected, nonuniform loads (Figure~\ref{fig:serverless-histograms}, left)
result in computation time distributions centered around a smaller mean and with
a more peaked shape compared to those of the uniform loads
(Figure~\ref{fig:serverless-histograms}, right). Because the delay time is
dominated by the computation time of workers in our experiments (\cf\
communication time in Figure~\ref{fig:serverless-histograms}), it also has a
similar behavior in its distribution. On the contrary, because the idle time is
a measure of the discrepancy between the fastest and slowest workers for a fixed
problem size and master-worker setup, it is decreased with uniform loads. As a
result, uniform loads result in less queuing times for workers. For instance, as
can be seen in Figure~\ref{fig:serverless-queue}, when we have $W = 256$
workers, workers with uniform loads still spend more time in computing than
idling, whereas those with nonuniform loads idle more. Unfortunately, having
workers spend more time computing than idling does not directly translate to a
more efficient algorithm, as ADMM iterations can sill converge to modest
accuracies with inexact $x$-minimization steps~\cite{2010-Boyd}.

\begin{figure}
  \centering
  \includegraphics{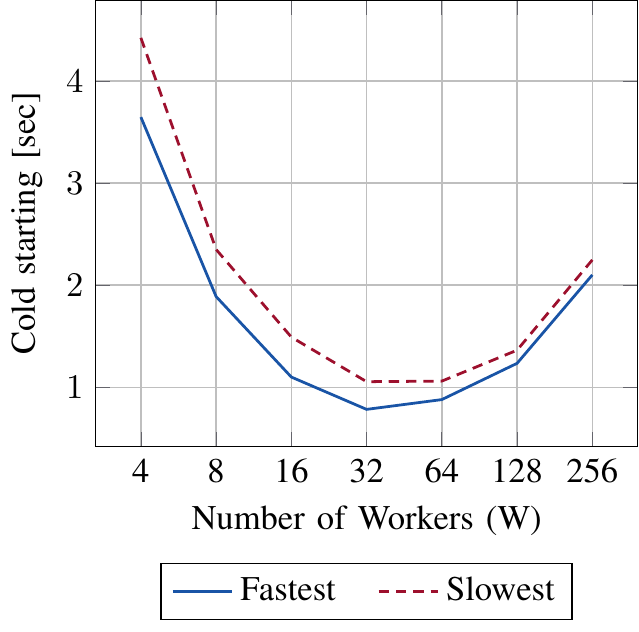}
  \caption{Cold starting of Lambda workers. The plot shows the first response
    time of the fastest and slowest workers after getting
    spawned.}\label{fig:serverless-cold-start}
\end{figure}

\textbf{Cold starting.} When generating AWS API Gateway requests, we use
\texttt{cURL}'s \emph{multi} interface that enables multiple simultaneous
transfers in the same (background) thread. We report cold-starting times of AWS
workers in Figure~\ref{fig:serverless-cold-start}, which is representative of
spawning new workers in \emph{bulks} with problem data that has closed form
representations. In the experiments, we observe that the cold starting of
workers are rather consistent, and, up to $W = 64$ workers, well below the
average time spent in computation per single ADMM iteration. Afterwards, the
cold starting degrades due to the queuing of bulk requests in the (background)
thread.

\begin{figure*}
  \centering
  \includegraphics{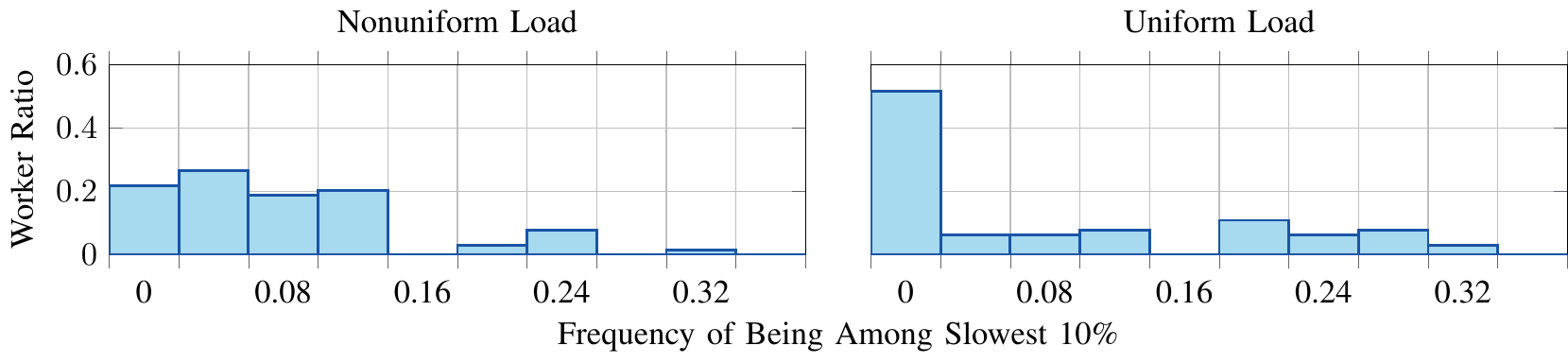}
  \caption{Responsiveness of Lambda workers during the algorithm when $W = 64$
    workers are used.}\label{fig:serverless-slowest}
\end{figure*}

\textbf{Responsiveness.} Finally, \replaced[id=MJ]{we compute the fraction of
iterations in which each worker is among the slowest 10\% to return its local
solutions to the scheduler, and plot the corresponding histogram in
Figure~\ref{fig:serverless-slowest}.}{ in Figure~\ref{fig:serverless-slowest},
we plot the ratio of workers that return their local solutions to the
scheduler in the slowest 10\% group.} Similar to the utilization metrics,
workers have consistent \deleted[id=MJ]{behaviour in their} responsiveness.
There are not any stragglers which fall behind more than one third of the
total iterations, and, only a very few of the workers lag behind
\replaced[id=MJ]{more than}{at least} one forth of the total time. Moreover,
the fastest group, \ie, the 0-bin in Figure~\ref{fig:serverless-slowest}, has
a bigger set of workers in uniform load scenario compared to that in the
nonuniform load.

\section{Conclusion}\label{sec:serverless-conclusion}

In this work, we have investigated the performance and limitations of serverless
runtimes when solving generic, distributed optimization problems. To this end,
we have built a master-worker setup in a star network, in which the central node
is a managed multi-core server and the other nodes are AWS Lambda functions. In
our experiments, we have used synchronous parallel ADMM iterations to solve
regularized logistic regression problems, and observed relative speedups up to
256 workers and efficiencies above 70\% up to 64 workers. Furthermore, even
though AWS Lambda does not give any specific performance guarantees, the workers
have satisfactory cold-starting times compared to their computation times and
they do not show major straggling problems that could hinder the performance of
the algorithms.

Because serverless runtimes are stateless and have limited compute times, they
have a major limitation when solving optimization problems. For long-lived
optimization algorithms, serverless runtimes require careful bookkeeping of
algorithm states as well as fault tolerance of workers approaching their time
limits. Second, inability to have inbound network connections at serverless
runtimes makes it impossible to use collective communication patterns such as,
\eg, MPI's \texttt{AllReduce} or \texttt{Bcast}, among the nodes.

\subsection{Outlook and Future Work}

Despite their aforementioned limitations, we believe that serverless runtimes,
with their availability and elasticity, are promising candidates for scaling the
performance of distributed optimization algorithms. There are some possible
algorithmic and system-level improvements to obtain better efficiencies, which
are left as future work.

\textbf{Algorithmic improvements.} In this work, we have considered a single
family of algorithms, \ie, synchronous parallel ADMM\@. We have observed that
increasing the computation times of worker nodes by making them use more
iterations in their local solvers does not directly translate to improved
efficiencies. One way to improve the parallel efficiency is to try asynchronous
parallel ADMM~\cite{2014-Zhang,2016-Chang,2016-Changa} or other (asynchronous)
families of algorithms that could potentially allow for better scalability. An
alternative approach could be to account for the slowest workers at each
iteration in the synchronized setting. In the machine learning community, there
have been recent works~\cite{2016-Venkataraman,2016-Chen,2018-Teng} that simply
discard a small percentage of the slowest workers in synchronized parallel
algorithms. In these works, discarding information contained within the slowest
workers' messages acts as an implicit regularization, and the authors obtain not
only improved timings but also better classification performance. However, for
generic optimization problems, this approach will result in a suboptimal
solution.  Instead, one can try coded optimization techniques~\cite{2017-Tandon,
2017-Karakusa,2017-Zhu} to alleviate the straggler effects in the synchronized
setting.

\textbf{System-level improvements.} In our experiments, we have solved problems
that involve decision vectors of size $d = 10\,000$. Broadcasting this vector
using point-to-point communications to workers, and reducing the information
coming from workers collectively using multiple masters have negligible effect
during computations (\cf\ communication and computation times in
Figure~\ref{fig:serverless-histograms}). However, for decision vectors with
sizes larger than, \eg, $d = 80\,000$, the communication time will be on par
with the computation time. In these cases, spawning masters as serverless
runtimes and using the ideas in~\cite{2018-Shankar} to replace the shared memory
of the masters with the high-bandwidth, high-latency distributed object store
could be beneficial in improving the communication times.

\bibliographystyle{IEEEtran}
\bibliography{IEEEabrv,library}

\end{document}